\documentclass[aps,amsfonts,twocolumn,pre]{revtex4}
\usepackage{epsfig,graphicx,amsmath,amsfonts,eucal,bm,mathrsfs}

\usepackage[latin1]{inputenc}
\newcommand{\beq}{\begin{equation}}
\newcommand{\eeq}{\end{equation}}
\newcommand{\bea}{\begin{eqnarray}}
\newcommand{\eea}{\end{eqnarray}}

\begin{document}
\title{On the Self-Affine Roughness of a Crack Front in Heterogeneous Media}
\author{Eran Bouchbinder, Michal Bregman and  Itamar Procaccia}
\affiliation{Dept. of Chemical Physics, The Weizmann Institute of Science, Rehovot 76100, Israel}

\begin{abstract}
The long-ranged elastic model, which is believed to describe the evolution of a self-affine rough
crack-front, is analyzed to linear and non-linear orders. It is shown that the nonlinear terms, while important in changing the front dynamics, are not changing the scaling exponent which characterizes the roughness of the front. The scaling exponent thus predicted by the model is much smaller than the one observed experimentally. The inevitable conclusion is that the gap between the results of experiments and the model that is supposed to describe them is too large, and some new physics has to be invoked for another model.
\end{abstract}
\maketitle

The self-affine roughness of a crack-front propagating under a tensile load in a randomly heterogeneous system is a well studied issue, both experimentally and theoretically. Experimentally one measures the
position $h(x,t)$ of the crack front, where $h$ is the position of the front as a function of the span-wise
coordinate $x$ at time $t$, and finds that this is a self-affine function whose
roughness is characterized by a scaling exponent $\zeta$ (defined below in Eq. (\ref{sscaling})) in the range of 0.5-0.65 \cite{crack, wetting}. Theoretically
there appears to be a consensus that the appropriate model for such dynamical roughening is
a long-ranged elastic string close to its depinning threshold. This model is defined by the
equation of motion for a front  $h(x,t)$, which is allowed to move only forward due to the
irreversibility  of the fracture process \cite{97REF}
\begin{eqnarray}
&&\!\!\!\!\!\!\!\!\frac{\partial h(x,t)}{\partial t} = G^{(0)}[1+2I_1]-\Gamma(x,h), \,{\rm for}~ \frac{\partial h(x,t)}{\partial t}\!>\!0, \label{lin}\\
&&I_1 \equiv\case {1}{2\pi}\int dx' \frac{h(x',t)-h(x,t)}{(x'-x)^2} \nonumber .
\end{eqnarray}
Here and bellow the integral is meant in the Cauchy Principal Value sense. The RHS of Eq. (\ref{lin})
is the difference between the local driving force
(below referred to as $G$, related physically to the energy release rate driving the crack \cite{Broberg}), and $\Gamma(x,h)$ which is a random quenched noise (representing the random material fracture energy \cite{Broberg}). $G^{(0)}$ is the control parameter that represents the energy release rate of
a straight front. The integral term stands for the long ranged restoring forces stemming from bulk elastic degrees of freedom. The correspondence between the theoretical model and the experimental findings remained however unclear, since the best numerical studies of the resulting self-affine graph $h(x)$ of this model came up with a roughness exponent $\zeta=0.388\pm0.002$ \cite{07DK}, clearly outside the range of error of the experimental measurements. This apparent difficulty led to a number of interesting studies, insisting that the model is basically right, and that the result concerning the scaling exponent is not final. Thus, for example, in \cite{01CL-DW} the authors analyzed Eq. (\ref{lin}) using a functional renormalization group. They have calculated the scaling exponent $\zeta$
to one- and two-loop orders in $\epsilon$, where $\epsilon=2-d$. To one-loop order the result is $\zeta=\epsilon/3$, predicting $\zeta=1/3$ at $d=1$,
deviating considerably from the best numerical estimate. To two-loop order the prediction increases to about 0.466, leading to a statement that the model probably describes properly the experimental findings. Unfortunately, it is well known that the $\epsilon$ expansion is often an asymptotic series \cite{01Z-J}, sometime providing a better estimate of the exponents at first order than at second order. For all that one knows the third loop-order may bring the exponent down, maybe even below the first loop-order. Another approach was advocated in Refs. \cite{06Ledoussal,06KA-B} who proposed that the discrepancy between model and experiment may be assigned to the
existence of nonlinear contributions to Eq. (\ref{lin}). In \cite{06KA-BV} the authors derived, in agreement with the results of \cite{Ramanathan}, that to second order in nonlinearity Eq. (\ref{lin}) reads \cite{foot2}:
\begin{eqnarray}
\frac{\partial h(x,t)}{\partial t} &=&G^{(0)}[1+2 I_1+I_1^2 +2I_2+\case{1}{4}h'^2]\label{nonlin}\\
&-&\Gamma(x,h)[1+\case{1}{2}h'^2]\ , \quad {\rm for}~ \frac{\partial h(x,t)}{\partial t}>0\nonumber\\
I_2\equiv \!\!\case{1}{4\pi^2}\!\! \int\!\!\! \int &&\!\!\!\!\!\!\!\!\!\!\!\!\frac{[h(x',t)-h(x,t)][h(x'',t)-h(x',t)]}{(x'-x)^2(x''-x')^2} dx'\! dx'', \nonumber
\end{eqnarray}
where the prime denotes a derivative with respect to $x$. Note that the coefficients are all determined
by elasticity theory and are not free. 
Using a method proposed by Schwartz and Edwards \cite{92SE}, it was
concluded that the nonlinear term affects the scaling exponent dramatically, stating that $\zeta\ge 0.5$ \cite{06KA-B}.
On the other hand in \cite{06Ledoussal} a similar nonlinear equation was analyzed in the
framework of one-loop renormalization group, yielding $\zeta\approx 0.45$. Clearly, the situation warrants some further scrutiny.

In this Letter we present careful numerical measurements of the scaling exponent of the present model to first and second order in $h(x,t)$. To this aim we simulate the dynamical model (\ref{lin}) with and without the nonlinear terms in Eq. (\ref{nonlin}). Our final conclusion is that although the second
order terms perturb the solution $h(x,t)$ significantly, they are actually irrelevant for the scaling exponent, that appears unchanged with or without the nonlinear terms. The uneasy conclusion of this analysis is that the model itself may not describe the experiment correctly; a discussion of this conclusion is offered at the end of this Letter.

To numerically simulate the model we discretize the spatial variable $x$, and swap
temporal changes with discretized steps in the
variable $h(x)$. Choosing the basic unit of length to be in the $x$ direction, we present below
simulations for $x\in[1,L]$ with $L=2^n$ in the range $[2048,16384]$. We used  periodic boundary conditions.  The discretization of $h(x)$ is chosen
in units of $1/7$; this seems arbitrary, but since in the depinning limit the velocity is irrelevant, this
discretization should not affect the scaling exponents. At step zero the interface is prepared with a random jitter to avoid spurious lattice artifacts. The random quenched noise $\Gamma(x,h)$, is picked at each lattice point from a uniform distribution in the interval $[0,1.5]$ \cite{foot}. Following \cite{98RF} we simulate the depinning limit
by increasing $G^{(0)}$ incrementally from zero, such that the local driving force at the least pinned
point overcomes the local fracture energy $\Gamma$. This local depinning may trigger additional motion until the interface stops, at which moment $G^{(0)}$ is increased further until the next weakly pinned point gives in. Measurements of the roughness
were taken when all the points $x$ moved at least one step after the last increment in $G^{(0)}$.

For the present calculation we employed the rms definition of the roughness, i.e.
\begin{equation}
w(\ell,L)\equiv\Bigg\langle \case {1}{\ell}  \sum_{x=j}^{j+\ell} \left[h(x)- \case{1}{\ell} \sum_{x=j}^{j+\ell} h(x)\right]^2\Bigg\rangle_j^{1/2} \quad \ell\le L
\end{equation}
We keep the implicit $L$ dependence in this quantity since it turns out that the roughness exponent is a slowly
convergent quantity as a function of $L$. Indeed, it was convincingly demonstrated in \cite{07DK} that the numerical value of roughness exponent as measured using the linear model (\ref{lin}) converges in the relation
\begin{equation}
w(\ell,L)\sim \ell^\zeta \ , \label{sscaling}
\end{equation}
only when $L$ is of the order of $10^6$. Not having simulations of this order we resort to finite size scaling
which was demonstrated \cite{Zhou} to  yield reliable exponents also with smaller values of $L$. The essence
of this method is the finite-size scaling assumption written as
\begin{equation}
w(\ell,L) = L^\zeta f(l/L) \ ,
\end{equation}
for $L$ very large there is a range of values of $\ell$ where $f(l/L)\sim (\ell/L)^\zeta$, coalescing with the
simple scaling assumption (\ref{sscaling}). For smaller values of $L$ one seeks the best value of $\zeta$
by data collapsing $w(\ell,L) L^{-\zeta}$ onto a universal function $f(\ell/L)$. An example of this procedure is shown in Fig. {\ref{lincollapse} in which the first order model had been employed and the results were averaged over 100 realizations for each $L$.
\begin{figure}
 \epsfig{width=.45\textwidth,file=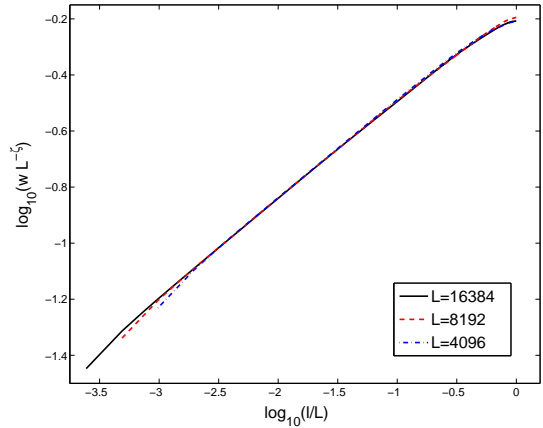}
\caption{Demonstration of data collapse for the linear model, plotting $\log_{10}(w(\ell,L) L^{-\zeta})$ vs. $\log_{10} (\ell/L)$ with $L=2^n$, n=12, 13 and 14, using $\zeta=0.362$. The results are obtained by averaging over 100 realizations.}
\label{lincollapse}
\end{figure}
The data collapse appears satisfactory with the choice $\zeta=0.362$. The degree of confidence that this
method provides can be demonstrated by the inferior data collapse obtained for the same data
with $\zeta=0.4$ in Fig. \ref{nocollapse}.
\begin{figure}
 \epsfig{width=.45\textwidth,file=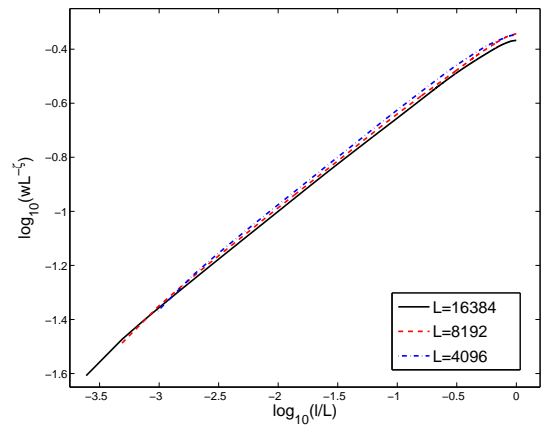}
\caption{Demonstration of the failure of data collapse for the linear model, plotting $\log_{10}(w(\ell,L) L^{-\zeta})$ vs. $\log_{10} (\ell/L)$ with $L=2^n$, n=12, 13 and 14, using $\zeta=0.4$. The results are obtained by averaging over 100 realizations.}
\label{nocollapse}
\end{figure}
Our best estimate of the scaling exponent of the model realized to first order is $\zeta=0.365\pm 0.005$.
Note that this exponent is higher than the value $\zeta\approx 0.35$ obtained
using Eq. (\ref{sscaling}) with $L=16384$. This is in agreement with the statements in the literature
for the slowness of convergence of the scaling exponent with $L$ \cite{07DK}. The finite size scaling analysis
improves the situation, even though our estimate still falls a bit short compared to the estimate $\zeta=0.388$ for $L\approx 10^6$ \cite{07DK}. This difference will not pose a difficulty in assessing the importance of the nonlinear term.

Adding the nonlinear terms, one should first ascertain that they make a significant change in the front dynamics. This is demonstrated in Fig. \ref{compare}, which compares, for the same initial interface and the same quenched noise $\Gamma(x,h(x))$, the realized interfaces, once with only a linear term and once with the full second-order nonlinear contributions.
\begin{figure}
 \epsfig{width=.45\textwidth,file=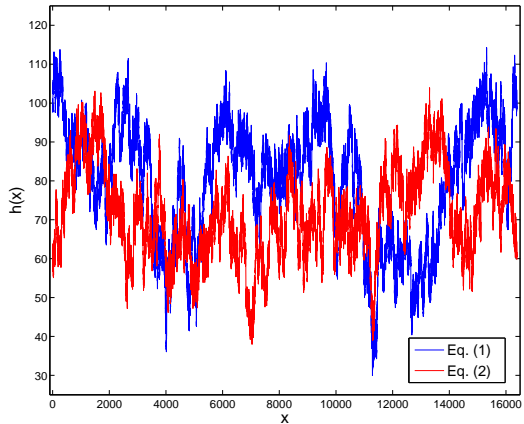}
\caption{Comparison of the fronts obtained with the linear and the nonlinear model, where the initial
condition and the quenched noise are all the same. Note the huge difference in scales between the abscissa and the ordinate.}
\label{compare}
\end{figure}
It is obvious that the nonlinear terms are not negligible, they inflict major changes on the actual
graph. The seasoned reader will notice however that the scaling exponent is hardly changed.
This eye-ball conclusion is fully supported by the finite-size scaling analysis which is presented in Fig. \ref{nonlincollapse}.
\begin{figure}
 \epsfig{width=.45\textwidth,file=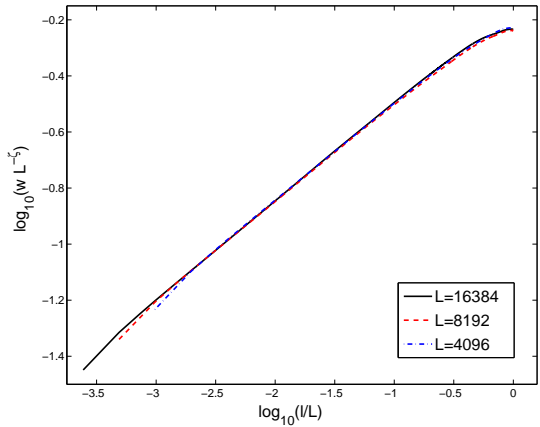}
\caption{Demonstration of data collapse for the nonlinear model, plotting $\log_{10}(w(\ell,L) L^{-\zeta})$ vs. $\log_{10} (\ell/L)$ with $L=2^n$, n=12, 13 and 14, using $\zeta=0.362$. The results are obtained by averaging over 100 realizations.}
\label{nonlincollapse}
\end{figure}
The quality of the data-collapse is essentially identical in Figs. \ref{lincollapse} and \ref{nonlincollapse}
{\em using the same scaling exponent in both cases}. Note that the amplitude of the overall front roughness is reduced in comparison to the linear model;  the nonlinearity increases the stiffness of
the front. The exponent remains however invariant.  We thus conclude that the numerical
evidence presented here does not support the theoretical propositions of \cite{06Ledoussal, 06KA-B}.

In light of these results we propose that the relation between the experiments and the model must
be reassessed. One could argue that our fronts are not large enough to asymptote to a ``correct" scaling behavior. To such a claim one must answer that the typical experiments do not have more scales
than our simulation. For example in the fracture experiments of \cite{crack} the randomness scale
(also known as the correlation length) is determined by the size of sand particles that blast the interface
between two slabs of material that are then glued together. This typical scale, which determines
the scale of the fracture energy $\Gamma$ in the present model, is of the order of 50$\mu$m. Crack front segment up to 50 mm were analyzed, giving at most three orders of scales in theory, but in practice the self affine scaling was observed in a range that spans about two orders of magnitude. The measured roughness exponent is significantly larger than the values discussed above, even in this limited range of length scales. It is therefore entirely reasonable, in our opinion, to restrict the theoretical analysis of any given model to about the same range of length scales or
slightly more, as is done above. Theories invoking asymptotically large system sizes may be interesting, but hardly relevant for such experiments.

Accordingly we may ask what is missing in the relation between theory and experiment. One thing that may be suspicious is the assumption at the background of the derivation of the models (\ref{lin})
and (\ref{nonlin}), i.e. that elasticity theory is entirely sufficient to describe the crack front dynamics.
Since elasticity theory predicts the divergence of stress at the crack front \cite{Broberg}, realistic materials
will almost surely yield plastically or develop additional local damage. Such a change in material properties, precisely where the dynamics
is taking place, may very well change the nature of the long-range interactions of the bulk degrees of
freedom. How to renormalize such long-range interactions when plasticity or other modes of damage
are at play is not known at this point in time. We stress, however, that the gap between the model results
and the experimental results indicate that such novel thinking about the theoretical fundamentals may
be unavoidable.

{\bf Acknowledgments}: 
We thank Pierre Le Doussal for reading the manuscript and for raising interesting questions.
This work has been supported in part by the Minerva Foundation, Munich, Germany, by the German -Israeli Foundation and by the Israel Science Foundation. E. B. is supported by the Horowitz Complexity Science Foundation.



\begin{thebibliography}{99}
\bibitem{foot2}
Note that the term proportional to $\Gamma(x,h) h'^2$ includes contribution to third order in
$h$ due to the implicit dependence of $\Gamma$ on $h$. We left this term as is since it is not
obvious how to single out the second order contributions.
\bibitem{foot}
There is no direct experimental information about the distribution of the quenched noise besides
its typical spatial scale (determined by the preparation with sand blasting, see \cite{crack}). The 
mean of the distribution affects the threshold for depinning.
\end{thebibliography}
\end{document}